# WASEF: Web Acceleration Solutions Evaluation Framework

MOUMENA CHAQFEH, NYUAD, UAE
RASHID TAHIR, University of Illinois Urbana-Champaign, USA
AYAZ REHMAN, LUMS, Pakistan
JESUTOFUNMI KUPOLUYI, NYUAD, UAE
SAAD ULLAH, LUMS, Pakistan
RUSSELL COKE, NYUAD, UAE
MUHAMMAD JUNAID, LUMS, Pakistan
MUHAMMAD ARHAM, LUMS, Pakistan
MARC WIGGERMAN, Vrije Universiteit Amsterdam, Netherlands
ABIJITH RADHAKRISHNAN, Vrije Universiteit Amsterdam, Netherlands
IVANO MALAVOLTA, Vrije Universiteit Amsterdam, Netherlands
FAREED ZAFFAR, LUMS, Pakistan
YASIR ZAKI, NYUAD, UAE

The World Wide Web has become increasingly complex in recent years. This complexity severely affects users in the developing regions due to slow cellular data connectivity and usage of low-end smartphone devices. Existing solutions to simplify the *Web* are generally evaluated using several different metrics and settings, which hinders the comparison of these solutions against each other. Hence, it is difficult to select the appropriate solution for a specific context and use case. This paper presents Wasef, a framework that uses a comprehensive set of timing, saving, and quality metrics to evaluate and compare different web complexity solutions in a reproducible manner and under realistic settings. The framework integrates a set of existing state-of-the-art solutions and facilitates the addition of newer solutions down the line. Wasef first creates a cache of web pages by crawling both landing and internal ones. Each page in the cache is then passed through a web complexity solution to generate an optimized version of the page. Finally, each optimized version is evaluated in a consistent manner using a uniform environment and metrics. We demonstrate how the framework can be used to compare and contrast the performance characteristics of different web complexity solutions under realistic conditions. We also show that the accessibility to pages in developing regions can be significantly improved, by evaluating the top 100 global pages in the developed world against the top 100 pages in the lowest 50 developing countries. Results show a significant difference in terms of complexity and a potential benefit for our framework in improving web accessibility in these countries.

## 1 INTRODUCTION

During the last decade, the technological ecosystem of the World Wide Web (WWW) has rapidly evolved to provide a richer and more engaging user experience through aesthetically appealing and highly-interactive pages. However, these advancements have resulted in significantly larger and more complex pages, with the median page size increasing by more than 80% in the last five years [39]. While the affordability of smartphone devices [3–8, 11] is driving the growth of mobile web browsing (68.1% of recent global web access is reported via mobile devices) [13], the web complexity is resulting in a poor browsing experience in developing countries due to slow connectivity and the low-end hardware of the affordable phone devices, popular in these countries [28]. A significant web performance degradation has been reported in Africa, South America, and India [21] due to the recent increase in traffic demand [30].

To remedy the poor browsing experience, several promising solutions have been proposed to manage the increasing complexity of the web [16, 23, 26, 31, 39, 45, 47, 48, 56, 62–64]. However, each of these solutions is evaluated with a different set of pages using different metrics, which makes it difficult to compare the solutions against each other, and hinders the developers' selection of the right solution according to their preferences and requirements of their expected pages' visitors. In addition, despite their numerous benefits, the lack of widespread access to these solutions and the missing unified evaluation of their effectiveness and applicability limit the ability to compare them against each other.

In light of these challenges, we present a framework for prototyping different scenarios for evaluating, comparing, and performance understanding of the benefits of web complexity solutions in a controlled experimental environment that enables reproducible results. Our framework combines the timing, energy saving, and quality metrics for a comprehensive evaluation with the consideration of multiple scenarios. The framework provides an extendable implementation that employs a set of existing solutions and allows for integrating other solutions in an easy manner. The framework crawls different types of pages, both landing and internal, and caches them to provide an evaluation baseline. For each web page, it runs each of these solutions separately to generate a specific version by each solution and cache it for a comparative analysis. We demonstrate how the framework can be used to evaluate the performance characteristics of different publicly available state-of-the-art web complexity solutions under realistic conditions. We analyze the variant benefits of these solutions on a sample dataset consisting of 1000 pages. We also show that pages accessed in developing countries tend to be more complex than the popular pages in the developed world, an observation that inspires a useful usage scenario for our framework to improve the web accessibility in these countries. In summary, we make the following contributions:

- Providing an evaluation framework considering different web complexity solutions, network conditions, and performance metrics.
- Maintaining a representative evaluation dataset consisting of both landing and internal pages.
- Analyzing the benefits of a set of publicly available web complexity solutions showing their performance gains impact on the structure and the functionality of both: landing and internal pages.
- Analyzing 100 popular web pages in the developing world in comparison to the the top 100 global popular pages, showing the difference in terms of complexity and the potential benefit from our framework.

## 2 MOTIVATION

Existing literature provides miscellaneous approaches to handle the increasing complexity of the web, including image compression [16, 39], page-load restructuring [47, 56, 63], and JavaScript optimization with non-essential elements removed [26] or unused functions eliminated [40, 45, 50]. Each of these approaches modifies a set of web components (such as images) or processes (such as the page load) in the original web pages to achieve better performing versions, and is evaluated with different metrics depending on the desired specific objective. Example objectives include but not limited to: accelerate the page load, save memory, reduce energy consumption, and accommodate the users data plans.

However, failing to show the impact of a web complexity solution on one of the key measurements might lead to incorrect assumptions about the benefits or misuse of the solution. For example, a memory-saving solution [45] might not achieve timing improvements, and thus, cannot be offered to users with connectivity issues, where page acceleration is crucial. Additionally, any optimizations to speed up the page load might impact the page content or interactive functionality [26]. Therefore, a clear picture of the benefits and the drawbacks would be only achieved by a comprehensive evaluation, such that the right solution can be used according to the user requirements and settings.

Conventional metrics can be categorized into three different classes: *timing*, *saving*, and *page quality* metrics. Another classification of these metrics can be found in [20]. In Table 1, we show the metrics used to evaluate the state-of-the-art web complexity solutions. As the table shows, most of these solutions choose either to measure the timing metrics or the energy consumption (memory, CPU, and Battery), while ignoring the quality evaluation in terms of content completeness and preservation of functionality. Nevertheless, few approaches [26, 39, 45] utilize small-scaled user studies to measure the quality of the generated pages. This can be explained by the challenges of conducting large-scaled user studies, mainly due to financial and/or logistic constraints. In the following, we discuss the main challenges in comparing web complexity solutions for performance analysis.

**Restricted access:** Most of today's web complexity solutions are not publicly available. Over time, even promising solutions might be abandoned by the community due to replication and application difficulties. One of the motivations behind this work is to create an atmosphere that encourages researchers to increase the accessibility to their solutions, by integrating them into a unified testing framework for a comparative, consistent, and standardized evaluation.

**Reproducibiliy**: Reproducing the results of evaluating a given web complexity solution is challenging even with the same list of pages, due to the rapidly growing complexity of WWW. Our proposed framework aims to "freeze" a large dataset of web pages for consistent and reproducible evaluation across different solutions. New instances of these pages can be constantly crawled to ensure a continuous representation of the web status, where a frozen version of each page is stored and the date of the data collection is reported to allow a later comparison to the same pages.

**Tools' limitations and rapid measurement changes** Different web complexity solutions are evaluated using different tools. Thus, a comparative analysis among these solutions might not be possible without further experiments. Additionally, none of the existing tools encompasses the three classes of the aforementioned measurements, and a complete picture of the benefits and drawbacks of a given solution cannot be drawn unless using a hodgepodge of individual tools. These challenges are further exacerbated by the fact that newer versions of existing tools are regularly released to provide improved measurements and cater to the fast-evolving WWW. For instance, Google Lighthouse 6 replaces a set of metrics that were available in the older version, with newer and more representative user-perceived timing metrics. Moreover, it introduces a new scoring algorithm that makes several amendments to the computations of the measurements considered by prior versions. A comparison between version 5 and 6 on the same set of pages reveals that 96% of the pages ended up with a different score [17]. This implies that solutions that were evaluated using Lighthouse 5 cannot be compared to those which were evaluated using Lighthouse 6 without further assessment.

**Evaluating the impact on the page's content & functionality:** Web complexity solutions often generate simplified versions from existing pages, where the original content and/or functionality might be sacrificed due to optimizations in resources and/or processes. Thus, a structural/functional assessment of the generated pages is crucial to ensure high-quality pages. However, this assessment is either ignored [31, 47, 48, 56, 63] or evaluated with small-scaled user studies [26, 39] due to the challenges of conducting large scale user studies and the lack of alternative evaluation scenarios.

## 3 RELATED WORK

### 3.1 Web Performance Measurement Tools

A number of tools are available for the evaluation of web pages' performance. WebPageTest [22] is a widely-used open-source web performance measurement and analysis tool with an online version available to the community, and sponsored by companies and individuals to provide testing agents infrastructure across the globe. It allows users to evaluate web pages from various locations and validate changes and audit performance. We believe that the features of WebPageTest make it an ideal environment to integrate and evaluate web complexity solutions and encourage their applicability in the developing world. In this work, we extend WebPageTest for that purpose. In addition to WebPageTest, there exist a set of publicly available tools that can be used to evaluate different aspects of the pages. For example, Lighthouse [12] is an open-source tool for performance measurements of web pages that runs a sequence of audits against a given page, and then generates a report that can be used to improve the page. Simiarly, a set of common web performance issues can also be reported using SEMrush [2] tool. Other commercial tools can measure web performance across different browsers, devices, and locations such as [14] and [10]. While the aforementioned tools provide efficient web performance evaluation environments, they mainly focus on the timing measurements of a single version of a given web page, and the consideration of different versions that can be offered by complexity solutions is still out of their current scope. These versions are usually evaluated with respect to the original version of a given web page with small-scaled user studies (See Section 2). To overcome the scaling and automation challenges of the user studies, the authors of Eyeorg [61] proposed a system for crowd-sourcing web quality measurements with controlled experimental conditions, which facilitates the evaluation of the impact of changes to the page structure with real users. In contrast, our framework relies on computer vision for larger scaled evaluation of the web contents structure. Additionally, the utilization of computer vision allows us to evaluate the functionality of web pages, a feature that is missing in Eyeorg. The evaluation of web pages in terms of energy consumption, CPU, and memory usage is fundamental for assessing mobile browsing. Android Runner [43] is a tool that can automatically execute experiments on Android devices to measure these metrics, which is integrated in our framework for that purpose. Other existing tools are either designed for a single measurement such as energy consumption, or ad-hoc for a specific experiment [54].

Table 1. Evaluation metrics of the state-of-the-art web complexity solutions (PLT is Page Load Time).

| Solution | Timing | Saving | Quality |
| --- | --- | --- | --- |
| WProf [62] | PLT | - | - |
| Flywheel [16] | PLT, Time-to-first-byte, Time-to-first-paint | Page Size | - |
| Shandian [63] | PLT | Page Size | - |
| Polaris [47] | PLT | - | - |
| Vroom [56] | PLT, Above-the-fold Time, SpeedIndex | - | - |
| Prophecy [48] | PLT, SpeedIndex ReadyIndex , | Bandwidth, Energy | - |
| SpeedReader [31] | PLT | Data Size, Memory | - |
| JSCleaner [26] | PLT, DOM Complete | Page Size | User Study |
| Web Medic [45] | - | Memory | User Study |
| BrowseLite [39] | Speed Index | Bandwidth, Page Size | User Study, Structure |

### 3.2 Web Measurement Studies

Existing web studies mainly focus on providing insights on the status of web tracking [29, 38, 41, 57], ad-blocking [36, 44, 58], and security and privacy [37, 51, 52, 60], with less focus on performance, accessibility, and user experience.

For example, web tracking with million pages is analyzed in [57] and in [29], with around 1500 pages in [38], and in a twenty-year period (from 1996 to 2016) in [41]. In [44], the impact of ad blockers is studied, whereas anti-ad-blocking is measured in [36], and the future of ad blocking is envisioned with an analytical framework in [58]. Additionally, an analysis of third-party requests is provided on a million websites in [42]. In the performance direction of web analysis, a recent study has assessed the performance of a set of 150,000 pages (both desktop and mobile) [59], and found that 82% of the analyzed pages have significant performance issues, while 44% of them had at least one critical issue. Such a study indicates that performance issues in modern web pages are common, and inspires the need for applying solutions to serve less fortunate users with low-end mobile web access. To understand the impact of different factors on improving mobile page load, another recent study [46] conducted over 4 years with 8 popular mobile browsers and 250,000 pages found that web pages are increasingly becoming more complex, and the observed improvements in mobile page performance is mainly due to the improved platform of mobile devices (not from page nor browser/network improvements). To the best of our knowledge, our analysis of 1000 web pages is the first to show the impact of web complexity solutions on different performance metrics, aiming towards applying these solutions at large scale.

## 4 WASEF FRAMEWORK

Wasef framework aims to overcome existing challenges in performance evaluation of web complexity solutions, by providing a platform for various solutions and measurements across Android handheld smartphone devices. In addition to the large consideration of Android in the literature [19], our selection of Andriod smartphones serves the objective of evaluating web complexity solutions across devices that are popular in developing countries [4–8]. Other core design considerations of our framework are:

- **Automation:** eliminating user interventions apart from the initial configuration and settings.
- **Extendability:** with a minimal integration effort, developers and researchers can integrate their own web complexity solutions in the framework and make them available via the user interface.
- **Replicability:** experiments are replicable with a given configuration, dataset, and smartphone device.
- **Comprehensiveness:** providing both timings and energy measurements, in order to assess the benefits and the drawbacks of each solution on both landing and internal pages.
- **Scalability of Assessment:** providing a fast automated assessment of the structure and the functionality of the pages created by web complexity solutions at a large scale, to present an alternative that overcomes the challenges and limitations of user studies.

### 4.1 Architecture

A high-level overview of the Wasef framework architecture is shown in Figure 1. Wasef consists of three main phases: processing (crawling web pages, running web complexity solutions on each page, and caching the optimized page versions generated by the solutions), evaluation (running measurement experiments), and results collection (to represent and output statistics from the conducted experiments).

In the processing phase, a set of web pages are crawled and cached in a dedicated proxy server. To provide a representative set of web pages, and due to the substantial differences between landing and internal pages in terms of content and structure as characterized in [18], both landing and internal pages are considered. These pages represent the original versions of the pages to be created through a set of integrated web complexity solutions. We selected the following web complexity solutions: Lacuna [50], JSCleaner [26], SlimWeb [25], Muzeel [40], and JSAnalyzer [24], based on the following criteria: 1) the availability of the source-code, 2) the potential generation of a simplified page from an original version, 3) the applicability on handheld mobile phone devices. By running these solutions on each original page, a simplified page is generated and cached at the proxy server. The crawled original pages along with their simplified versions provide a dataset for evaluation under various configurations. This evaluation consists of three main experimental-based measurements: timing (page load plus a set of user perceived timing metrics), energy (battery consumption, CPU, and memory usage), and structural/functional assessment of the simplified pages with respect to their original versions. To the best of our knowledge, these three main measurements cover a comprehensive set of evaluation metrics considered by today's web community, which are not yet available together in a unified framework.

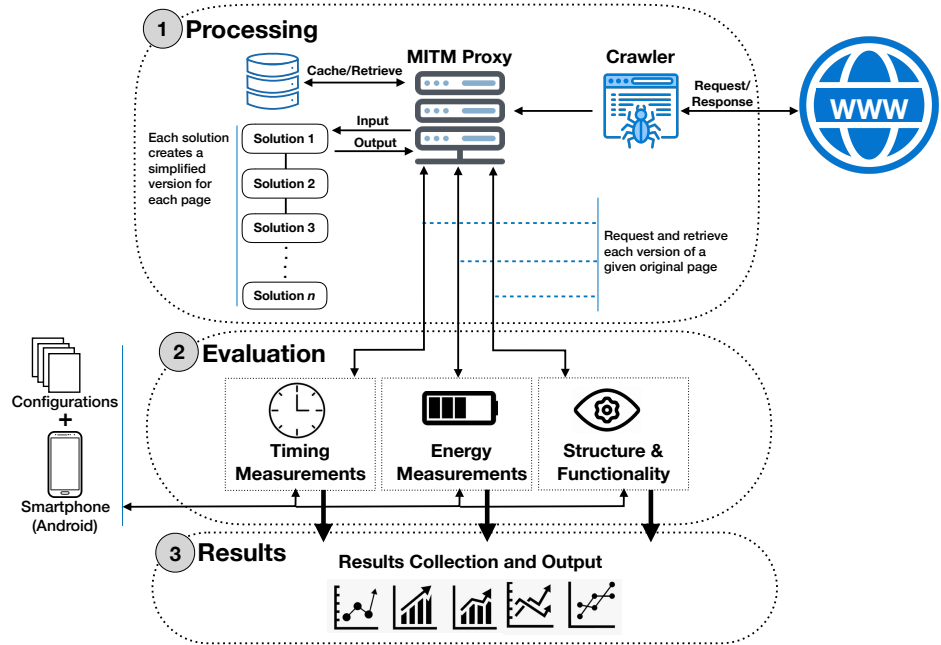

Fig. 1. Wasef Framework Architecture

### 4.2 Performance Measurements

Our framework considers three groups of performance metrics for a comprehensive comparative analysis; timing, energy, and page structure/functional quality metrics.

*4.2.1 Timing Measurements.* The classical timing metric was the page load time (PLT), which does not adequately capture the experience of real users. Instead, timing metrics such as time-to-interactivity [49], and speed index [34] have instead been proposed to measure the user-perceived performance of the page load. Our framework considers all the timing metrics of WebPageTest [22] spanning from the first contentful paint to the full page load time. In addition to the timing metrics, bandwidth consumption measurements are also reported including the page download size and the number of network requests.

*4.2.2 Energy Measurements.* These measurements include battery, CPU, and memory consumption, which are collected via the Android utilities as described in [43].

*4.2.3 Structure and Functionality Assessment.* Two metrics are considered to assess the structure and the functional quality of the pages generated by the web complexity solutions with respect to their original versions, which are: 1) *Structural Similarity*, which is a score given to the simplified page to assess the content completeness including text, images, and interactive components with respect to the corresponding original version, and 2) *Functional Similarity*, which is a score given to the simplified page to evaluate the completeness of functionality in comparison to the original page. This assessment is driven by computer vision as an alternative to large-scale user studies, where the structure of a given simplified page is assessed by a score of similarity to the corresponding original page, and the functionality of that page is evaluated by emulating the user interactivity with both the simplified and the original page to return a score of functionality completeness.

## 4.3 Implementation

We extended an existing implementation of MITM proxy [27] along with Selenium [35] and Chrome driver to crawl web pages from the web and cache them in a dedicated proxy server. To integrate each of the web complexity solutions in our framework, we extend the WebPageTest [22] tool to allow the selection of these solutions via the user interface of the tool. To integrate the energy measurements, we extend the AndroidRunner [43] framework that automatically executes experiments on Android devices to measure battery consumption, CPU, and memory usage. Additionally, we integrated Qlue [32, 33] to evaluate both the structural and the functional similarity of the pages created by the employed complexity solutions with respect to their original versions. Web developers and researchers can extend the Wasef framework in two ways. First, by integrating their web complexity solutions to allow for the generation of new set of simplified pages. Second, by integrating their own measurement tools into the framework via the extension point provided by the Plugin handler [43] of AndriodRunner.

## 5 EVALUATION

Our evaluation aims to show the benefits of Wasef in creating a unified environment for evaluating different web complexity solutions, and to provide a comparative analysis to assess each of the employed solutions from the perspectives of timing, power, page structure and functionality preservation. Our evaluation consists of two experiments, 1) a comparative analysis of a set of publicly available web complexity solutions, where each solution is used to generate simplified versions of a set of 1000 original web pages, and 2) an analysis of the performance of the top 100 globally popular pages versus a set of 100 pages popular in developing countries. A typical Wasef experiment involves a server running the software modules for conducting a sequence of experiments to assess a set of web pages including the original page and the simplified versions derived from the employed solutions.

### 5.1 Experimental Setup

We dedicate a powerful server machine with 64 cores and 1 TB of RAM to crawl and cache a set of 1000 popular web pages including both landing and internal pages, as recently recommended by Hispar [18]. We selected 500 pages from the Hispar list [55] and another 500 from Tranco [15, 53] (which lists popular web pages from Alexa, Cisco Umbrella, and Majestic). Starting from the first page in each set, we filtered out the pages with a screenshot of size less than 1MB to exclude lightweight pages. The selected 1000 pages are loaded via the Chrome browser while recording the full

HTTP(S) content and headers using an extended version of MITM proxy [27]. The server also hosts the Wasef software modules to: 1) run each of the employed web complexity solutions on the cached pages and store the output version generated by each solution, 2) conduct a comprehensive evaluation for each version by considering all the measurements sequentially, starting from the timing metrics, followed by the energy consumption, and then the structure/functionality computations. For a given original web page, all the versions are generated automatically apart from JSAnalyzer pages that are created by a number of recruited web developers and then cached by the server. To evaluate the structural similarity of the pages with respect to their original versions, we monitored the screenshot generation to ensure high-quality screenshots, where elements that repeatedly appear while scrolling (such as floating banners) are removed.

Additionally, to create a dataset for the second experiment, we crawled the top 100 globally popular pages from the Tranco's top million sites in 2021. From the same list, we also crawled the top 100 sites found from the lowest 50 countries in the list of developing countries across the globe, sorted based on the human development index [1].

To represent a slow cellular network in both experiments, we configured Wasef to emulate a 3G network with a bandwidth of 1.6 Mbps and 400 milliseconds of Round Trip Time (RTT). We selected Xiamoi Redmi Go as a representative low-end Android smartphone device (OS: Android 8.1 Oreo (Go edition), CPU: Quad-core 1.4 GHz Cortex-A53, RAM: 1GB, Battery: Li-Ion 3000 mAh), which connects to the Internet over a fast WiFi (with upload and download bandwidth of 100 Mbps). Pages are automatically requested by firing the Chrome browser on the phone, which is the most popular browser [9].

**JSAnalyzer Pages' Generation:** *JSAnalyer* was evaluated by 10 users, who were invited to generate simplified pages using JSAnalyzer through social media groups by introducing the tool and providing a contact email. Received requests were accepted in a first-come-first-served basis. Each user was invited for an online training session to analyze a web page. Users were able to select an hour time-slot from a set of available times/dates. In each session, a user was introduced to the purpose of JSAnalyzer, and then had a quick usage tutorial on how to use it. Users were given access to install JSAnalyzer to analyze existing web pages and produce new simplified pages. An online evaluation session ends when a user evaluates a single page successfully and saves the simplified version. The time spent on each session depends on the complexity of the page being analyzed (5-30 minutes). By the end of an online evaluation session, the user is asked to continue analyzing a set of 15 pages offline, ending up with a total of 150 pages analyzed by 10 users.

### 5.2 Web Complexity Solutions Analysis

In this section, we evaluated five of the different state-of-the-art web complexity solutions—the ones that had publicly published their code—against each others. The evaluations were structured to compare these solutions from three different high-level performance angles, namely:

- Timing metrics: highlighting how quickly the pages' elements were loaded.
- Benchmarking metrics: highlighting the different resources utilizied by the pages.
- Qualitative metrics: highlighting the pages' structural and functional similarity to their original counter-parts.

Additionally, each solution was evaluated over two different sets of pages: 500 pages from the Tranco-list (representing the top accessed landing pages around the world), and 500 pages from the Hispar-list (representing the top accessed internal pages around the world).

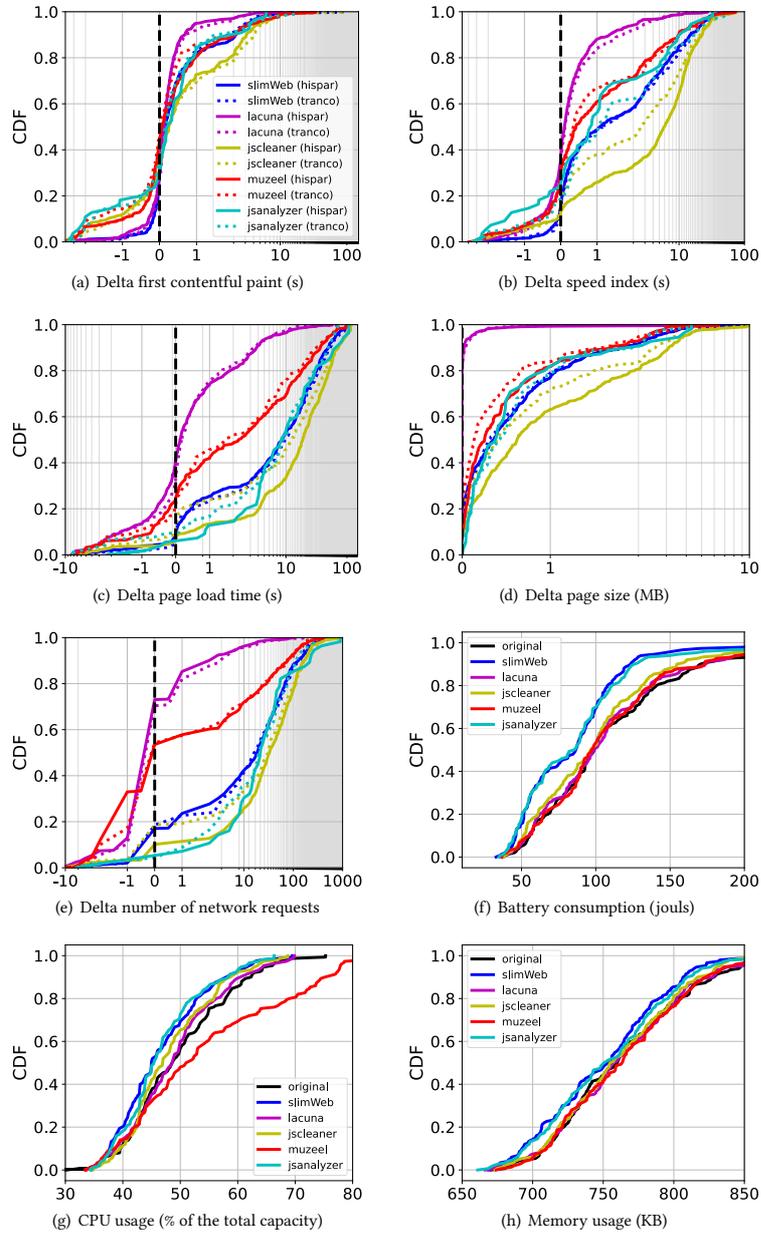

Fig. 2. Quantitative comparison of web complexity solutions

**Timing metrics:** Figures 2(a) to 2(e) show the Cumulative Distribution Functions (CDFs) of the delta performance results of each page version (named with the solution used to generate it) with respect to the corresponding original page. The delta is computed by subtracting the value of a given metric measured for a specific version of the page from the value measured for the original page. Figure 2(a) shows the delta first contentful paint metric, which is a timing metric referring to how quickly the first visual elements get displayed on the browser. It can be seen from the figure that *JSCleaner* (represented by the yellow curve) has the highest delta across the different solutions. On the other hand, *Lacuna* (represented by the magenta curve) shows the lowest delta gains across all solutions. The figure also shows that there is very marginal difference of the solutions when used over landing pages vs. internal ones. Next, the speed index metrics results are shown in Figure 2(b). Speed index is a popular timing metric that is used to measures how quickly content is visually displayed during page load. It can be seen from the figure again that *JSCleaner* has a much higher delta speed index compared to all other solutions, roughly about six seconds faster median speed index when compared to the original pages. Additionally, *JSCleaner* seems to improve the internal pages' speed index better than the landing pages. Figure 2(c) shows the page load time CDF results of all solutions. In here, the results highlight that both *slimWeb* and *JSAalyzer* have slightly lower delta page load times when compared to *JSCleaner*. On the other hand, *Lacuna* shows a very moderate delta page load times when compared to all other solutions. Similar trends can be observed for both the delta page size (Figure 2(d)) and the delta number of network requests (Figure 2(e)).

**Benchmarking metrics:** To study the impact of the web complexity solutions on the mobile phone utilized resources, we compare three different metrics against each other: "Battery consumption", "CPU utilization", and "Memory usage". Figures 2(f) to 2(h) show the CDFs of the reported measurements for all the versions in addition the original, respectively. Analyzing the battery consumption results show that the soultions basically split into two different groups, where the first group consists of both *slimWeb* and *JSAnylzer*, whereas the second group has the rest of the solutions. The first group achieves a much lower battery consumption compared to the rest of the solutions, suggesting that both *slimWeb* and *JSAnalyzer* lower down the battery consumption due to their blocking of non-essential JavaScript. For the other two benchmarking metrics, CPU and memory, it can be observed that no clear differences can be noticed among the solutions apart from *Muzeel* that seems to consume more CPU in about 60% of the pages when compared to the other solutions.

**Qualitative metrics:** No analysis of web complexity solution is complete without assessing the impact of such a solution on the page visual content and functionality, i.e., assessing the content and functional similarity to the original page. Hence, Wasef utilized Qlue to compute the structural and functional similarity of all the pages generated by the above five solutions in comparison to the original pages, as depicted by Figure 3.

Figure 3(a) shows the CDFs of structural similarity, where it can be seen that *JSCleaner* has the lowest similarity scores in comparison to all other solutions, achieving on median about 80% structural similarity score, in comparison to all other solutions that are maintaining above 95% similarity scores. Observing the functional similarity scores in Figure 3(b) shows that, apart from *Muzeel*, all solutions have a median functional similarity scores of about 85%. In contrast, *Muzeel* retains more than 95% functional similarity scores compared to the original.

**Concluding remarks:** To conclude the results of the web complexity solutions comparison, we need to evaluate all of the metrics together. Although the timing metrics revealed that *JSCleaner* outperforms all other solutions in terms of the first contentful paint, speed index, the page load time, page size, and the number of network requests. However, these savings come with an impact on the page structure and (see Figure 3(a)) and functionality (see Figure 3(b)). In contrast, *Lacuna* shows the least performance gains in terms of the metrics evaluated in Figures 2(a) to 2(e), but it presents the best structural similarity score (see Figure 3(a)). On the other hand, the impact of *Lacuna* on the pages' functionality can be explained by the lack of user interactivity evaluation after the page loads, resulting in marking the interactivity functions as *unused code*. This limitation has been resolved in *Muzeel* (which also relies on dead-code elimination), where improved performance is shown in the functional similarity (see Figure 3(b)).

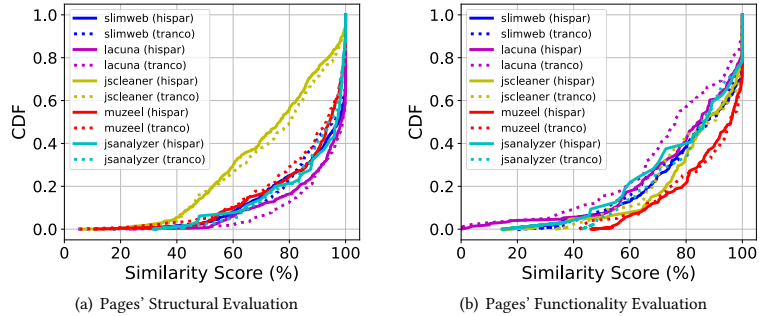

(a) Pages' Structural Evaluation  (b) Pages' Functionality Evaluation

Fig. 3. Qualitative similarity comparison

These two examples above (*JSCleaner* and *Lacuna*) give a clear evidence that evaluating web complexity solutions on a subset of metrics is not enough to draw a clear conclusion. This additionally highlights the benefits of using a framework like Wasef, where a complete analysis can be performed from different angles to study the impact of such a solution.

Interestingly, *JSAnalyzer* significantly reduces the page load time and the number of network requests in comparison to *SlimWeb*. In terms of CPU usage, *Muzeel* shows the worst CPU utilization among all solutions. It can also be seen that blocking solutions such as, *slimWeb*, and *JSAnalyzer* have generally a lower battery consumption in comparison to the other solutions. Finally, no significant impact can be seen in terms of the memory consumption for all the different solutions.

### 5.3 Web Pages Complexity: Developing vs. Developed

In this section used the Wasef framework to assess the difference between the top 100 pages from the developing regions and the top 100 global pages from the developed world. The motivation behind this analysis is to understand the differences between developed world pages in comparison to developing regions' pages, in terms of pages' complexity and the potential of web complexity solutions on such pages. We crawled the tranco-list top 100 global pages from 2021, as well as the top 100 pages from 2021 of the lower one-third developing regions sorted based on their human development index [1]. We have chosen *slimWeb* as the web complexity solution to be evaluated on these two sets of pages, in order to quantify its speedup impact on the pages. Given that *slimWeb* operates by blocking non-essential JavaScript in order to speedup the page load time, we evaluated the impact on the overall JavaScript processing time of the browser, as well as the overall page load time.

---

[1]This was taken from the full list of 152 developing regions reported in https://www.worlddata.info/developing-countries.php.

Figure 4(a) shows the CDF of the JavaScript processing time for both the developing regions and the developed world pages. The red curves represent the CDFs of the developing regions' pages, where the solid curve is the result of the original pages, whereas the dashed curve represents the results of the *slimWeb* solution. Similarly, the developed world pages' results are shown in green, with the sold curve referring to the original and the dashed curve representing *slimWeb*. The results show that developing regions' pages requires higher JavaScript processing time by the mobile phone browser in comparison to the developed world pages. This is due to the additional inefficiencies when designing and building pages in developing regions, where developers tend to miss-use the implementations by adding many unnecessary JavaScript libraries and files. Figure 4(b) shows the average JavaScript processing time for both pages in addition to the *slimWeb* pages, with the error bars represent the 95% confidence interval. The results show that the average time spent on processing JavaScript code is about 10 seconds for the developing regions' pages in comparison to about 7 seconds for the developed world pages. One can also observe that the error bars of the developing regions tend to be larger than the developed world suggesting a higher fluctuations in the processing time across the developing regions' pages. Finally, the figure also show that using a solution like *slimWeb*, depicted by the dotted hatches, does indeed reduce the processing time significantly for both pages, reducing the processing time by about 40% in comparison to the original version.

The page load time CDF and average results are shown in Figure 4(c) and 4(d) respectively. Similar to the JavaScript processing time results, the page load time for the developing regions' pages is higher than the developed world ones. However, the benefits of using a solution like *slimWeb* seems to be slightly higher for the developed world pages than the developing regions' ones, achieving a reduction of 30% for the developed world pages in comparison to a 22% reduction for the developing regions' pages.

## 6 DISCUSSION AND CONCLUSION

While the development of solutions to tackle the web complexity has shown significant improvements in accelerating the page load and saving the data plans, the lack of a unified framework for a comparative analysis hinders the applicability of these solutions. Our framework provides a rapid and consistent evaluation of web complexity solutions, to take an important step towards facilitating decisions on the selection of solutions in developing regions according to the users' needs and preferences, envisioning three use case scenarios: 1) researchers aiming to analyze the performance of their proposed solutions against existing approaches to provide a full picture on the benefits and drawbacks from different perspectives. 2) developers aiming to decide on a solution (or a combination of solutions) to accommodate the requirements of both the context of the page (content-rich, highly-interactive, reader-mode) and the different target user groups. 3) In case of multiple versions being provided by the developer for a given web page, users can be served with the version of the page that accommodates their settings (network condition and device) and preferences (speeding up the page load or saving bandwidth). Our analysis of five versions of 1000 web pages generated by existing web complexity solutions spanning different categories demonstrate that significant speed ups in the page load can increase the CPU consumption and impact the page structure and interactivity.

For the future, we plan on combining page components' optimizations with processing speed-ups to address the challenge of maximizing timing and energy gains while maintaining the page structure and interactive functionality. Additionally, we plan to open-source Wasef, and provide an online testbed with several servers, storage units, and different types of phones for the research community to submit and run customized experiments, and integrate their own web complexity solutions for evaluation. We hope that Wasef would contribute to the improvement of the browsing experience in developing countries, and to the provision of a better understanding of the pages' complexity in these countries.

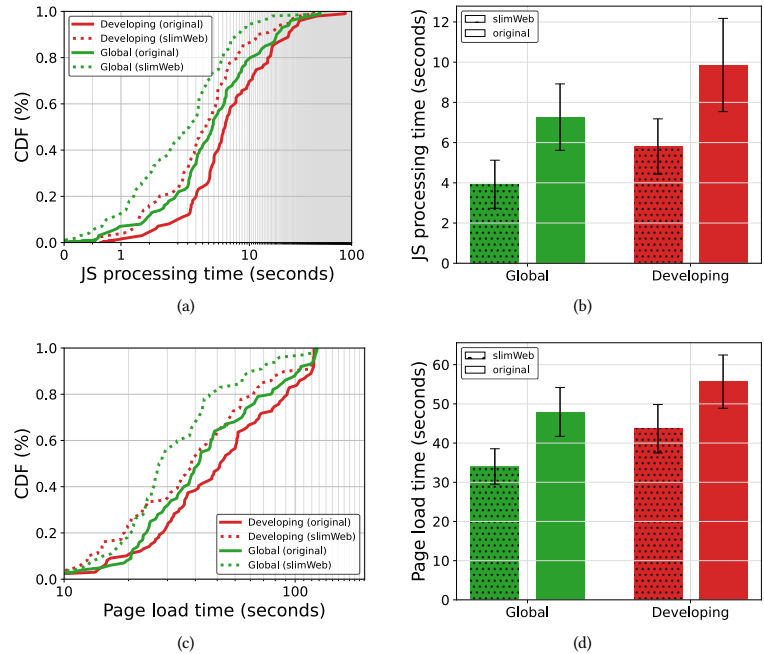

Fig. 4. Q